\documentclass[11pt]{amsart}

\title{ Linear decomposition attack on   public key exchange protocols using semidirect products of (semi)groups}

\author{Vitali\u\i\ Roman'kov}
\address{Institute of Mathematics and Information Technologies\\Omsk State Dostoevskii University}
\curraddr{}
\email{romankov48@mail.ru}

\usepackage{amsmath,amsthm,amsfonts,amssymb}
\usepackage{graphicx}
\usepackage{hyperref}
\usepackage[usenames]{color}

\theoremstyle{definition}

\newcounter{comcount}

\def\MA{{\mathbf{A}}}
\def\F{{\mathbb{F}}}

\date{}

\begin{document}

\maketitle

\begin{abstract}
 We show that a linear decomposition attack based on the decomposition method introduced by the author in  monography \cite{R1} and  paper \cite{R2} works  by finding the exchanging keys in the both two main protocols in \cite{HKKS} and  \cite{KLS}. 
\end{abstract}

\section{Introduction}
\label{se:intro}

In this paper we present a new practical  attack on   two main protocols proposed in \cite{HKKS} and \cite{KLS}.  This kind of attack  introduced by the author in \cite{R1} and \cite{R2} works when the platform  groups  are linear. We show that in this case, contrary to the common opinion (and some explicitly stated security assumptions), one does not need to solve the underlying algorithmic problems to break the scheme, i.e.,  there is another  algorithm that recovers the private keys without solving the principal algorithmic problem on which the  security assumptions are based. This changes completely our understanding of security of these scheme. The efficacy of the attack depends on the platform group, so it requires a  specific analysis in each particular case. In general one can only state that the attack is in polynomial time in the size of the data, when the platform and related groups are given together with their linear representations. In many other cases we can effectively use known linear presentations of the groups under consideration. A theoretical base for the decomposition method is described in \cite{RM} where a series of examples is presented. The monography \cite{R1} solves uniformly  protocols based on the conjugacy search problem (Ko, Lee et. al. \cite{KL}, Wang, Cao et. al \cite{WC}), protocols based on the decomposition and factorization problems (Stickel \cite{S}, Alvares, Martinez et. al. \cite{AM}, Shpilrain, Ushakov \cite{SU}, Romanczuk, Ustimenko \cite{RU}), protocols based on actions by automorphisms (Mahalanobis \cite{M}, Rososhek \cite{Ros}, Markov, Mikhalev et. al. \cite{MM}), and a number of other protocols. See also \cite{R3} where the linear decomposition attack is applied to the two main protocols in \cite{WXLLW}.

In \cite{KLS}, D. Kahrobaei, H.T. Lam and V. Shpilrain described a public key exchange protocol based on an extension of a  semigroup  by automorphisms (more generally endomorphisms). They proposed a non-commutative semigroup of matrices over a Galois field as platform. 

In this paper we present a polynomial time deterministic attack that breakes the two variants of the protocol presented in the papers  \cite{HKKS} and \cite{KLS}. 

All along the paper we denote by $\mathbb{N}$ the set of all positive integers.

\section{General key exchange protocol \cite{HKKS}, \cite{KLS}.}

In this section, we describe a not platform-specific key exchange protocol proposed in \cite{HKKS} and improved in  \cite{KLS}. We consider the more general version of this protocol presented in \cite{KLS}. The corresponding version from \cite{HKKS} has been analyzed in \cite{RM}. Then we will give a cryptanalysis of this protocol under additional assumption of linearity of the chosen platform. 

Let   $G$ be a (semi)group and $g$ be a public element in $G.$ Let $\phi $ be an arbitrary public endomorphism  of  $G.$ Let $G_{\phi }=G\  \lambda \ sgp(\phi )$ be the semidirect product of $G$ and the semigroup $sgp(\phi )$ generated by $\phi .$ Recall that each element of $G_{\phi}$ has a unique expression of the form $(\phi^{r}, f)$ where $r \in \mathbb{N}\cup \{0\}$ and $f \in G.$ Two elements of this form are multiplied as follows:  $(\phi^r, f)\cdot (\phi^s, h) = 
(\phi^{r+s}, \phi^{s}(f)h).$

\begin{itemize}
\item Alice chooses a private $m \in \mathbb{N},$ while Bob chooses a private $n \in \mathbb{N}.$ 
\item Alice computes $(\phi , g)^m = (\phi^m, \phi^{m-1}(g) \cdots \phi^2(g)\cdot \phi (g)\cdot g)$ and sends only the second component  $a_m = \phi^{m-1}(g) \cdots \phi^2(g)\cdot \phi (g)\cdot g$ of this pair to Bob. 
\item Bob computes $(\phi , g)^n = (\phi^n, \phi^{n-1}(g) \cdots \phi^2(g)\cdot \phi (g)\cdot g)$ and sends only the second component  $a_n = \phi^{n-1}(g) \cdots \phi^2(g)\cdot \phi (g)\cdot g$ of this pair to Alice.
\item Alice computes $(\ast , a_n)(\phi^m, a_m) = (\ast , \phi^m(a_n)a_m).$ She does not actually "compute" the first component of the pair. 
\item Bob computes $(\ast , a_m)(\phi^n, a_n) = (\ast , \phi^n(a_m)a_n).$ He does not actually "compute" the first component of the pair.  
\item Since $\phi^m(a_n)a_m=\phi^n(a_m)a_n=a_{m+n},$ we should have $K_{Alice} = K_{Bob} = a_{m+n},$ the shared secret key. 
\end{itemize}

This algorithm can be named the {\it noncommutative shift}.

Now we show how the shared secret key $K = K_{Alice} = K_{Bob}$ can be computed in the case when $G$ is a multiplicative subgroup of a finite dimensional algebra $\MA$  over a field $\mathbb{F}$ and  the endomorphism $\phi $  is extended to an endomorphism of the underlying vector space $V$ of $\MA$.  Furthermore, we assume that the basic field operations in $\mathbb{F}$ are efficient, in particular they can be performed in polynomial time in the size of the elements, e.g., $\mathbb{F}$ is finite. In all the particular protocols considered in this paper  the field $\mathbb{F}$ satisfies all  these conditions.

Using Gauss elimination we can effectively find  a maximal linearly independent subset $L$ of the set $\{a_0, a_1, ..., a_k, ... \},$   where $a_0 = g$ and $a_k = \phi^{k-1}(g) \cdot ... \cdot \phi (g) \cdot g$  for $k \geq 1.$ Indeed, suppose that $\{a_0, ..., a_k\}$ is linearly independent set but 
$a_{k+1}$ can be presented as a linear combination of the form

$$a_{k+1} = \sum_{i=0}^k\lambda_ia_i, \  \textrm{for} \   \lambda_i \in \F.
$$

Suppose by induction that $a_{k+j}$ can be presented as above for every $j \leq t-1.$ In particular

$$a_{k+t-1} = \sum_{i=0}^k\mu_i a_i,  \  \textrm{for } \  \mu_i \in \F.$$

Then

$$a_{k+t} = \phi (a_{k + t-1}) \cdot g =    \sum_{i=0}^k\mu_i \phi (a_i) \cdot g =$$
$$\sum_{i= 0}^{k}\mu_i a_{i+1} =  \mu_k\lambda_0a_0 + \sum_{i=0}^{k-1}(\mu_i + \mu_k\lambda_{i+1})a_{i+1}.$$
Thus $L = \{a_0, ..., a_k\}.$  

In particular, we can effectively compute 

\begin{equation}
\label{eq:1}
a_n = \sum_{i=0}^k \eta_i a_i, \   \textrm{for } \  \eta_i \in \F.
\end{equation}

Then 
$$
a_{m+n} = \phi^m(a_n)\cdot  a_m = 
$$
\begin{equation}
\label{eq:2}
\sum_{i=0}^k \eta_i \phi^m(a_i) \cdot a_m
= \sum_{i=0}^k \eta_i \phi^i(a_m) \cdot a_i.
\end{equation}

Note that all data on the right hand side of (\ref{eq:2}) is known now. Thus we get the shared key $K = a_{m+n}.$

In the original version of this cryptosystem  \cite{HKKS}    $G$ was proposed to be the  semigroup of $3 \times 3$ matrices over the group algebra $\F_7[\mathbb{A}_5]$,
where $\mathbb{A}_5$ is the alternating group on $5$ elements. The authors of \cite{HKKS}  used an extension of the semigroup $G$ by an inner automorphism which is conjugation by a matrix $H \in $ GL$_3(\F_7[\mathbb{A}_5]).$  Therefore, in this case   there is a polynomial time algorithm to find the shared key $K$ from the public data.

\section{Key exchange protocol using matrices over a Galois field and extensions by special endomorphisms \cite{KLS}.}

In this section, we describe the key exchange protocol using matrices over a Galois field and extensions by special endomorphisms  proposed in \cite{KLS}. 

Let $G$ be a multiplicative  semigroup of the matrix algebra $\MA = $ M$_2(\mathbb{F})$ of all $2 \times 2$ matrices over the Galois field $\mathbb{F}= \mathbb{F}_{2^{127}}.$ Let $\varphi = \sigma_H$ be the automorphism of  $G$ which is a composition of a conjugation by a matrix $H \in $ GL$_2(\mathbb{F})$ with the endomorphism $\psi $ that raises each entry   
of a given matrix to the power of $4.$ The composition is such that $\psi $ is applied first, followed by conjugation. Note that both these maps naturally  extend to automorphisms of $\MA .$

This protocol can be attacked by the linear decomposition attack as it has been explained in Section 2.

In \cite{KLS}, the situation is considered where the  automorphism $\varphi $ is just conjugation by a public matrix $H\in $ GL$_2(\mathbb{F})$. Let $g = M \in G.$ By direct computation one get $a_k = H^{-k}(HM)^k$ for every $k \in \mathbb{N}.$ 

This protocol is vulnerable to a linear algebra attack as follows. The attacker, Eve, is looking for matrices $X$ and $Y$ such that $XH=HX$, $Y(HM)=(HM)Y$, and 
$XY = H^{-m}(HM)^m.$  The first two matrix equations translate into a system of linear equations in the entries of $X$ and $Y$ over $\mathbb{F}.$ After solving this system and finding invertible solution $X$ and $Y$, Eve can recover the shared secret key $K$ as  follows: $Xa_nY = H^{-n}(XY)(HM)^n = H^{-n}H^{-m} (HM)^m(HM)^n = H^{-(m+n)}(HM)^{m+n} = a_{m+n}=K.$ The above algorithm contains a couple of difficulties. Firstly, a solution $X$ might be invertible. Secondly, all this computations should be done online during every session. 

In contrast to  the linear algebra attack, the linear decomposition attack is very simple. We describe even a more simple version of this attack working in this specific situation. 

Consider the linear space $W=$ Sp$_{\mathbb{F}}(gp(H)\cdot sgp(HM))$ generated by all elements of the form $H^k(HM)^l$ where $k, l \in \mathbb{N}\cup \{0\}.$ One can find effectively a basis $e_1, ..., e_t$ of $W.$ Obiously, $t \leq 4.$ Moreover, since every matrix is a root of a characteristic polynomial of degree $2$  one can choose  basic elements in the form $e_i = H^{k_i}(HM)^{l_i}, k_i, l_i \in \{0, 1\}, i = 1, ..., t.$ 
Now we have public dates $a_m$ and $a_n$ where $m, n \in \mathbb{N}.$ We can effectively compute 

\begin{equation}
\label{eq:3}
a_n = \sum_{i=1}^t \eta_i e_i= \sum_{i=1}^t\eta_i H^{-k_i}(HM)^{l_i}, \   \textrm{for } \  \eta_i \in \F, i = 1, ..., t.
\end{equation}

Then 

$$
\sum_{i=1}^t \eta_i H^{-k_i}a_m(HM)^{l_i}=\sum_{i=1}^t \eta_i H^{-k_i}(H^{-m}(HM)^m) (HM)^{l_i}=$$
$$ =H^{-m}(\sum_{i=1}^t \eta_i H^{-k_i}(HM)^{l_i})(HM)^m=$$

\begin{equation}
\label{eq:4} =H^{-m}H^{-n}(HM)^n(HM)^m = H^{-(m+n)}(HM)^{m+n} = a_{m+n}.
\end{equation}

Thus one has the shared key $K = a_{m+n}.$ Note that the basis $e_1, ..., e_t$ is constructed one time offline. We don't need to look in any invertible solution. 

In \cite{KLS}, the last protocol was changed to avoid the linear algebra attack. As before $H, M \in G,$  where $H$ is invertible and $M$ is assumed to be  not invertible.  The automorphism $\varphi $ is $\sigma_H,$ the inner automorphism corresponding to $H.$ 

\begin{itemize}
\item Alice chooses a private $m \in \mathbb{N},$ while Bob chooses a private $n \in \mathbb{N}.$ Alice also selects a private nonzero matrix $R$ such that $R\cdot (HM) = 0$ (the zero matrix), and Bob selects a private nonzero matrix $S$ such that $S \cdot (HM) = 0.$ Such matrices $R, S$ exist because the matrix $HM$ is not invertible. 
\item Alice  computes $(\varphi , M)^m = (\varphi^m, \varphi^{m-1}(M) \cdots \varphi^2(M)\cdot \varphi (M)\cdot M)$ where the second component of this pair is    $a_m = \varphi^{m-1}(M) \cdots \varphi^2(M)\cdot \varphi (M)\cdot M=H^{-m}(HM)^m,$ and sends $a_m+R$ to Bob. 
\item Bob computes $(\varphi , M)^n = (\varphi^n, \varphi^{n-1}(M) \cdots \varphi^2(M)\cdot \varphi (M)\cdot M)$, where the second component  is $a_n = \varphi^{n-1}(M) \cdots \varphi^2(M)\cdot \varphi (M)\cdot M=H^{-n}(HM)^n,$ and sends $a_n + S$ to Alice.
\item Alice computes $(\ast , a_n+S)(\varphi^m, a_m) = (\ast , \varphi^m(a_n+S)a_m).$ She does not actually "compute" the first component of the pair. She only needs the second component of the pair, which is $H^{-(m+n)}(HM)^{m+n} + (H^{-m}SH^m)\cdot (H^{-m}(HM)^m).$ Since $S\cdot (HM) = 0,$ so Alice gets $K_{Alice} = a_{m+n}.$
\item Bob computes $(\ast , a_m+R)(\varphi^n, a_n) = (\ast , \varphi^n(a_m+R)a_n).$ He does not actually "compute" the first component of the pair. Similarly, he gets $K_{Bob}= a_{m+n}.$ 
\item Alice and Bob have the shared secret key   $K=K_{Alice} = K_{Bob} = a_{m+n}.$  
\end{itemize}

It is shown in \cite{KLS} that the linear algebra attack as above does not work against this protocol. Unfortunately, this protocol is vulnerable against the linear decomposition attack as follows.

Consider the linear space $W$  generated by all elements of the form $H^{-k}(HM)^k$ where $k = 1, 2, ... .$ Note that $a_m, a_n \in W.$ Let $U$ be the annihilator space of $HM$ consisting of all matrices $A \in \MA $ such that $A\cdot (HM) = 0.$ Note that $R, S\in U.$ Let $Z = W + U.$ 
One can find effectively a basis $e_1, ..., e_l, f_1, ..., f_t$ of $Z,$ where $e_i \in W, i = 1, ..., l;$ $f_j \in U, j = 1, ..., t.$  Let  $e_i = H^{-k_i}(HM)^{k_i},$ where $  k_i \in \mathbb{N}, i = 1, ..., l.$ 

Now we have public dates $a_m+R$ and $a_n+S$ where $m, n \in \mathbb{N},$ and we know that $ R, S \in U.$ We can effectively compute
\begin{equation}
\label{eq:5}
a_n+S  = \sum_{i=1}^l \eta_i e_i +  \sum_{j=1}^t \nu_j f_j = 
\sum_{i=1}^l\eta_i (H^{-k_i}(HM)^{k_i} + S_1, 
\end{equation}

\noindent   where   $ \  \eta_i, \nu_j \in \F$ for $ i = 1, ..., l$ and $ j = 1, ..., t,$  and  $S_1 \in U.$ It is possible that $S_1 \not= S.$

Then 
$$
\sum_{i=1}^l \eta_i H^{-k_i}(H^{-m}(HM)^m+R) (HA)^{k_i}=$$
$$ =H^{-m}(\sum_{i=1}^l \eta_i H^{-k_i}(HA)^{k_i})(HM)^m=$$

\begin{equation}
\label{eq:6} H^{-m}(H^{-n}(HM)^n-S_1)(HM)^m = H^{-(m+n)}(HM)^{m+n} = a_{m+n}.
\end{equation}

Thus one has the shared secret key $K = a_{m+n}.$ Note: 1) the basis $e_1, ..., e_l, f_1, ..., f_t$ is constructed one time offline, 2) we don't need to look in invertible solution of considered sets of linear equations along the algorithm works.  We apply the usual Gauss elimination process to find unique solution every time when we solve sets of linear equations in the algorithm.  Hence, this algorithm is deterministic.  Moreover, in the case where the platform is such or similar as proposed in \cite{KLS} the algorithm is practical. Note:  we don't compute $m$ and/or $n$ to recover $K.$

\end{document}